\documentclass[final]{svjour3}
\usepackage{graphicx}
\usepackage{rotating}
\usepackage{amssymb}
\usepackage{siunitx}
\usepackage{array}
\newcolumntype{M}[1]{>{\centering\arraybackslash}m{#1}}
\DeclareSIUnit\sq{\ensuremath{\Box}}
\usepackage{mathptmx}
\usepackage[numbers,sort]{natbib}

\makeatletter

\journalname{Journal of Low Temperature Physics}
\bibpunct{[}{]}{,}{n}{}{,}

\begin{document}

\newcommand{\hdblarrow}{H\makebox[0.9ex][l]{$\downdownarrows$}-}
\title{Prototype high angular resolution LEKIDs for NIKA2}

\author{S. Shu$^{1,\dagger}$ \and M. Calvo$^{2,3}$  \and S. Leclercq$^1$ \and J. Goupy$^{2,3}$ \and A. Monfardini$^{2,3}$ \and E.F.C. Driessen$^1$}
%\institute{IRAM}
%\author{sss}

\institute{$^1$ Institut de RadioAstronomie Millim\'{e}trique, 300 rue de la Piscine, 38406 Saint-Martin d$'$H\'{e}res, France\\
$^2$ Universit\'{e} Grenoble Alpes, 621 avenue Centrale, 38400 Saint-Martin-d$'$H\`{e}res, France\\
$^3$ Institut N\'{e}el, CNRS, BP 166, 38042 Grenoble, France\\
$^\dagger$\email{shu@iram.fr}}

\maketitle

\begin{abstract}
The current resolution of the NIKA2 \SI{260}{\GHz} arrays is limited by the $1.6\times$\SI{1.5}{mm^2} inductor size on the individual pixels. In view of future updates of the instrument, we have developed a prototype array with smaller pixels that is experimentally compared to the current pixel design. In-lab we find an increase of the angular resolution of \SI{8}{\percent}, promising an on-sky FWHM resolution of $10.2\si{\arcsecond}$ using this new pixel design.


\end{abstract}

\section{Introduction} 

The kinetic inductance detector (KID) has been a promising detection technology for astronomical observations since its first publication in 2003~\cite{Day:2003a,Zmuidzinas:2012a,Farrah:2017a}. Several projects like ARCONS~\cite{Mazin:2013a}, NIKA~\cite{Monfardini:2010a} and NIKA2~\cite{Adam:2018a} using lumped element KIDs (LEKIDs)~\cite{Doyle:2008a} have successfully achieved astronomical observation results.

\begin{figure}
\centering
\includegraphics[scale=0.33,trim={1cm 8cm 0.8cm 8cm},clip]{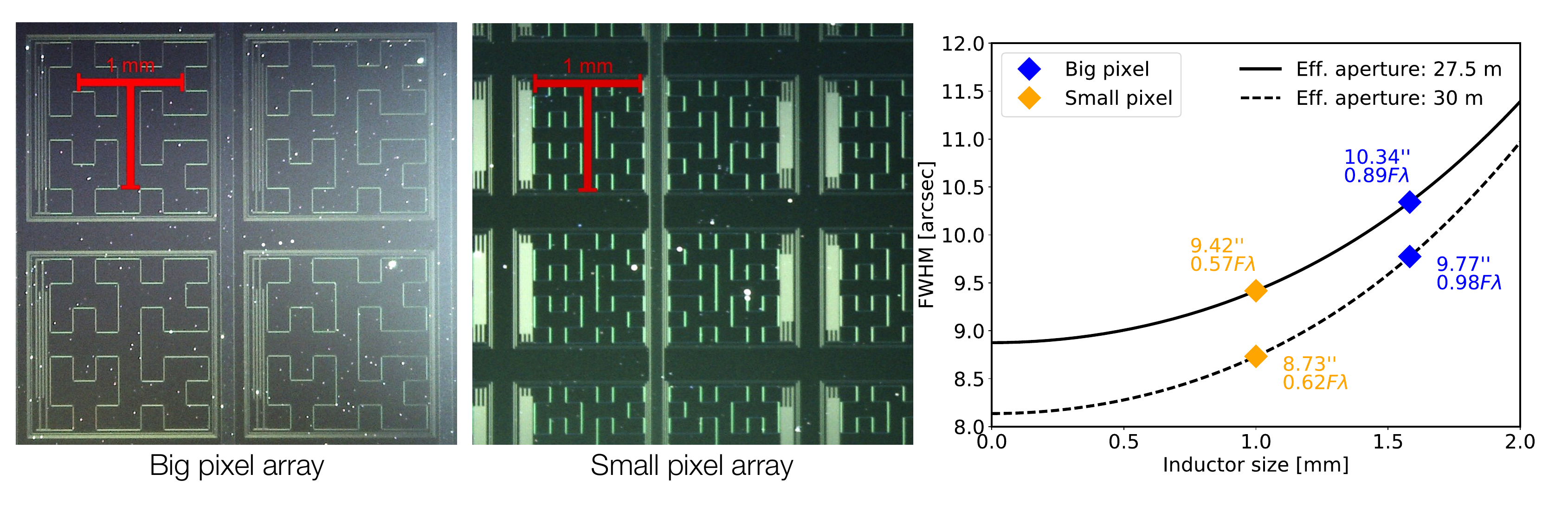}
\caption{\label{fig:res} {\it Left} two panels are microscopic images of the current NIKA2 \SI{260}{\GHz} design and our small pixel design. The red bars indicate \SI{1}{mm}. {\it Right} panel shows the calculated resolution as a function of inductor size, for \SI{27.5}{m} (unvignetted part of the primary mirror defined by the cold pupil) and \SI{30}{m} (full illumination of the primary mirror) effective telescope apertures.}
\end{figure}

The New IRAM KID Array 2 (NIKA2) is a LEKID-based millimeter-wave instrument dedicated to the IRAM 30-m telescope operating for millimeter astronomy~\cite{Calvo:2016a,Adam:2018a}. It consists of one 616-pixel array for the \SI{150}{\GHz} band and two 1140-pixel arrays for the \SI{260}{\GHz} band. The current NIKA2 \SI{260}{\GHz} array is designed to fully cover the \SI{80}{\milli\metre} diameter focal plane (6.5 arc-minutes field-of-view) with a large effective absorption area. With limited readout lines and bandwidths, an inductor size of \SI{1.6}{\milli\metre} was designed to achieve $10.5\si{\arcsecond}$ full-width half-maximum (FWHM). NIKA2 has been deployed and commissioned from October 2015 to April 2017 at the telescope. During the commission campaigns, an average $10.9\si{\arcsecond}$ FWHM on sky resolution is successfully achieved~\cite{Adam:2018a}. However, the advantage of the high angular resolution (theoretically $8.9\si{\arcsecond}$ FWHM) of the 30-m telescope is not fully developed. Increasing the angular resolution allows the instrument to reach the confusion limit, defined by the telescope aperture. In this paper, we present a prototype array design (small pixel) to achieve a higher angular resolution in the \SI{260}{\GHz} band. We experimentally characterize this prototype and compare it to the currently deployed NIKA2 pixels (big pixel). For comparison, Fig.~\ref{fig:res} shows microscope images of the two pixels studied in this paper, taken at the same magnification.
Decreasing the pixel size also decreases the inductor volume, which increases the responsivity. Also, more pixel can be placed on the focal plane. In total, the total performance of the NIKA2 instrument will be improved.

\section{Array Design}

The resolution of a detector is determined by the size of the effective beam resulting from the convolution of the diffraction pattern, created by the instrument optics, with the pixel transfer function. LEKIDs are intrinsically multi-mode detectors where the sensitive area is situated on the inductors which are also radiation absorbers. There are two ways to improve the resolution of a multimode pixel. One is to increase the $F\lambda$ by revising instrument optics; the other is to decrease the pixel size. Since $\lambda$ is the observation wavelength, a constant in design, increasing $F\lambda$ results in a large f-number $F$ which would need an increase of the focal plane size. To predict the resolution by changing pixel size, first we calculate the convolution of the telescope Airy pattern and the pixel response function which is assumed as to be a square of the size of the inductor~\cite{Griffin:2002a}. Secondly, the resolution is fitted from the convoluted result with a Gaussian function. The results are shown in Fig.~\ref{fig:res}.

Unlike horned detectors for which the illumination of the primary mirror is apodized, bare pixel arrays require the use of a cold pupil vignetting the edge of the primary mirror in order to minimize spill over from a warm background (diffraction on the mirror edge). For the current NIKA2 optics, the effective aperture size, \SI{30}{m}, is decreased by the cold pupil to \SI{27.5}{m} which gives $F\lambda = \SI{1.77}{mm}$. For the prototype design, we decrease the inductor size from \SI{1.6}{mm} to \SI{1}{mm}. With this small pixel design, theoretically a \SI{9.4}{\arcsecond} FWHM angular resolution is expected, which is close to the diffraction limit \SI{8.9}{\arcsecond} FWHM, as shown in Fig.~\ref{fig:res}. This resolution actually is achieved at the band center frequency of \SI{260}{GHz} (\SI{1.15}{mm} wavelength) and the actually achieved resolution will be decreased by the long wavelength part in the band. Decreasing the pixel size also decreases the effective area if the pixel number is constant. To keep full focal plane coverage, a doubled number of pixels is necessary while decreasing the pitch distance from \SI{2}{mm} to \SI{1.4}{mm}. For the future development of NIKA2 system, an increased readout bandwidth with \SI{1000}{\MHz} is under development. 

For the new pixel design we kept the order $3$ Hilbert inductor design~\cite{Roesch:2012a}. The inductor width is scaled from \SI{4}{\micro\metre} to \SI{2.5}{\micro\metre} for keeping the same effective sheet impedance of the inductor part while decreasing inductor length. With the same film thickness, the inductor volume of the small pixels is \SI{40}{\percent} of that of the big pixels. The Al film is assumed to be \SI{20}{nm} thick with transition temperature $T_c=\SI{1.4}{\kelvin}$, sheet inductance $L_s=\SI[per-mode=symbol]{2}{\pico\henry\per\sq}$, and sheet resistance $R_s=\SI[per-mode=symbol]{1.6}{\ohm\per\sq}$. For the Hilbert curve, we have 63 line segments with \SI{145}{\micro\metre} length. The interdigital capacitors are designed based on the readout \SI{500}{MHz} bandwidth and the final resonances are tuned from \SIrange{2.121}{2.621}{GHz} with simulations in Sonnet~\cite{Sonnet,Wisbey:2014a}. The coupling quality factor $Q_c$ to the \SI{120}{\micro\metre} width microstrip feedline is designed from \numrange{1.3e4}{1.6e4} to match the internal quality factor under typical background illumination at the IRAM 30-m telescope.

\begin{figure}
\centering
\includegraphics[scale=0.165,trim={1cm 8cm 3cm 8cm},clip]{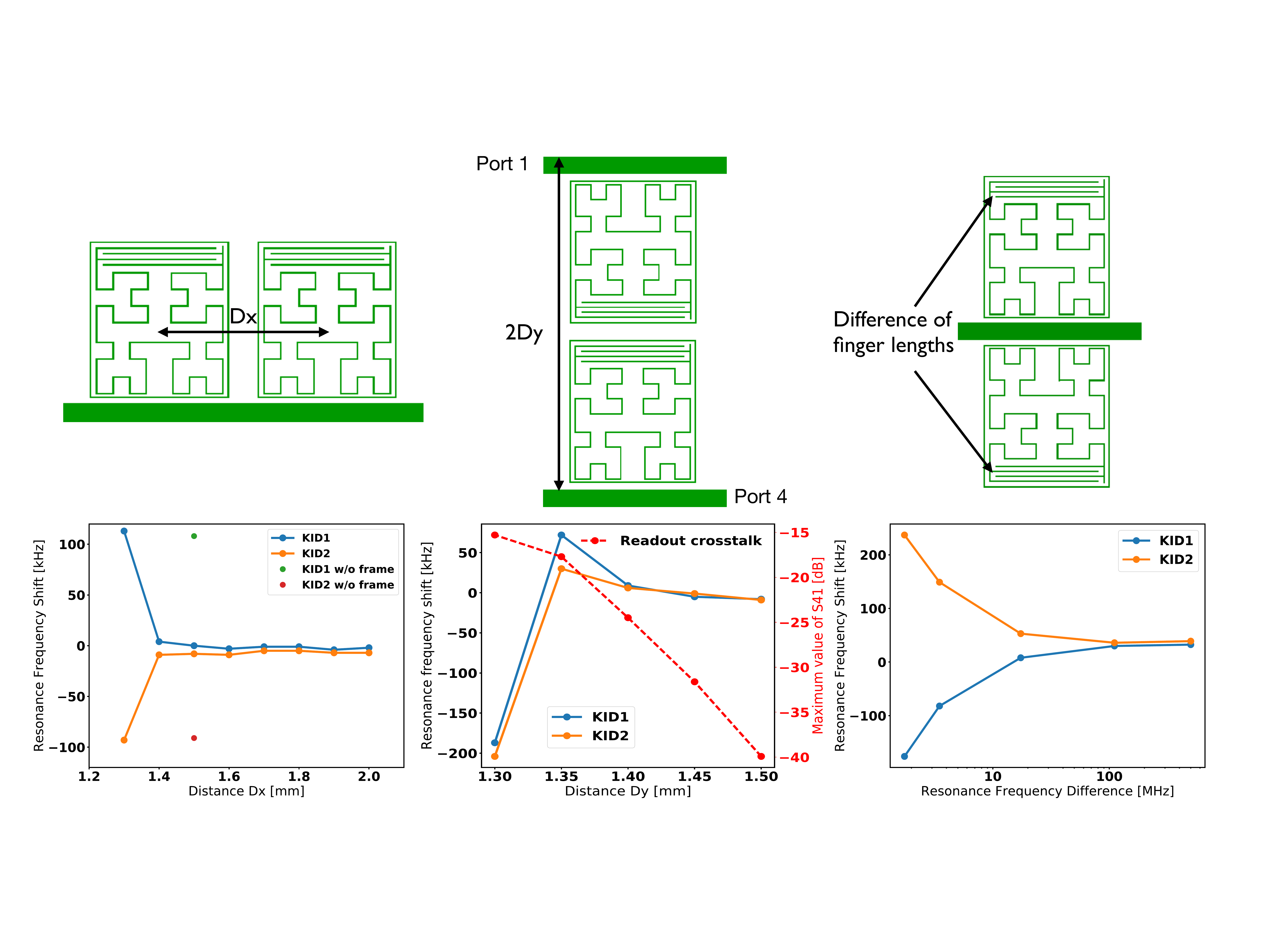}
\caption{\label{fig:xtalk}Simulated crosstalks in three different configurations. {\it Left} panel shows the crosstalk with different horizontal pitch distance Dx. Two single {\it solid symbols} show the crosstalk without the frame outside the LEKID. {\it Central} panel shows the crosstalk with different vertical pitch distance Dy and the maximum transmitted energy between two readout line. {\it Right} panel shows the crosstalk with different resonance frequency by changing the capacitor finger length.}
\end{figure}

The focal plane layout of bare LEKIDs is mainly limited by the microwave crosstalk caused by the electromagnetic coupling between nearby resonances~\cite{Noroozian:2012a}. This crosstalk is quantified as the difference between single LEKID resonance frequency and the resonance frequency of the same LEKID simulated together with another LEKID in a specific configuration shown in Fig.~\ref{fig:xtalk}. Three highest crosstalk configurations are simulated to find a most compact layout for this small pixel. Beyond a knee of \SI{20}{\kHz} shift of the intrinsic resonances the crosstalk becomes significant and increases extremely fast with decreasing distance between neighbor pixels. As seen in our simulation (Fig.~\ref{fig:xtalk}), to get as close as possible to this knee, the pitch between pixels along readout line Dx is set to \SI{1.4}{mm} and pitch between two readout lines $2\times \textrm{Dy}$ is set to $2\times \SI{1.4}{mm}$. To decrease the effective surface occupied by the readout line, pixels are placed on both sides of the feedline, in contrast with the pixel design currently deployed in NIKA2. This design causes the crosstalk to become significant when the two resonances are too close to each other either in frequency or spatially. To minimize this crosstalk, we tune the frequency difference of two opposite pixels to the maximum frequency separation, \SI{250}{\MHz} for \SI{500}{\MHz} readout bandwidth. This decreases the crosstalk to a constant \SI{35}{\kHz} offset for both resonances.

\section{Measurements}
\subsection{Fabrication}
For this study, we have produced two arrays in the IRAM cleanroom. To allow a thorough comparison, the arrays were produced using exactly the same fabrication procedure. First, a layer of \SI{20}{\nm} Al film is deposited onto a high-resistivity ($>\SI{5}{\kilo\ohm\centi\metre}$) \SI{250}{\um} thick silicon substrate using electron-beam evaporation. On the backside of the substrate, a layer of \SI{200}{\nm} Al film is deposited as a backshort using DC magnetron sputtering. This thick backshort is designed to reflect incident wave losslessly by impedance mismatch. Contact lithography is used for front-side Al, followed by wet etching to pattern the design in the Al film.

\subsection{Optical measurements}

\begin{figure}
\centering
\includegraphics[scale=0.3,trim={3cm 9cm 3cm 9cm},clip]{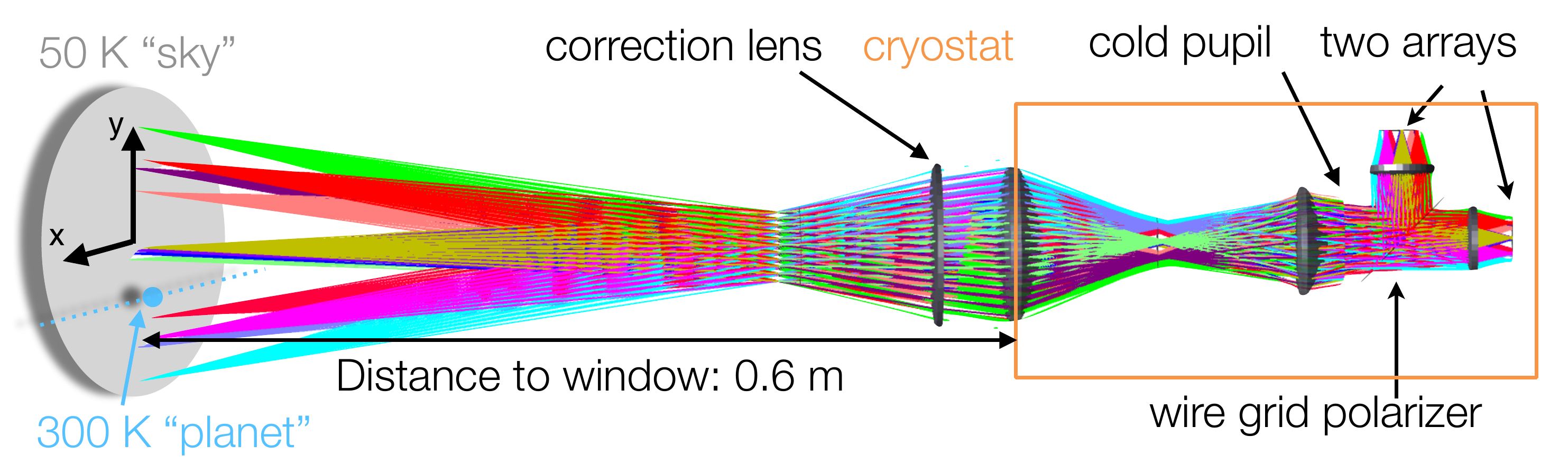}
\caption{\label{fig:optics}Optical setup of the experiment. The sky simulator is located \SI{0.6}{m} away from the cryostat window and the planet is scanned in the plane parallel to this window. Inside the cryostat two arrays are mounted on the \SI{70}{mK} stage. The incident signal is split by a wire grid polarizer, and both arrays are oriented identically with respect to the incident polarization. With this setup, the image size of the planet on the two arrays is the same as the image size of a point source on the focal plane of NIKA2 at the IRAM 30-m telescope.}
\end{figure}

A sky simulator and cold optics setup with $F=1.48$ are used to characterize the angular resolutions of both designs, shown in Fig.~\ref{fig:optics}. The sky simulator consists of a high temperature point like source mounted in front of a low temperature background plate. The point like source is a \SI{4}{mm} diameter metal ball (planet) at \SI{300}{\kelvin}, which is mounted with a transparent thin nylon line. The planet is placed at a distance of \SI{60}{cm} to the entrance window of the cryostat containing the optics and the detectors, and a \SI{50}{\kelvin} blackbody, cooled down by a pulse tube cryocooler, is behind the planet acting as a sky background~\cite{Roesch:2012a}. The two detector arrays are mounted on the \SI{70}{mK} stage of the cryostat (which uses a $^3\textrm{He}$-$^4\textrm{He}$ dilution fridge), in the focal plane of the optics after a polarizer, which allows us to measure both arrays simultaneously, and reduces measurement errors induced by the setup. The \SI{260}{\GHz} band is defined by a \SI{10.15}{cm^{-1}} low pass filter and a \SI{5.65}{cm^{-1}} high pass filter\footnotemark and the incident polarization directions are parallel to the capacitor fingers for both arrays to maximize the optical absorption. With this setup, the image size of the planet on focal plane is the same as the image size of a point source on the 30-m telescope focal plane.

\footnotetext{Both filters are from Cardiff University}

During the measurements, an x-y table moves the planet along the y-axis, as shown in Fig.~\ref{fig:optics}, to perform a single scan. By changing the x-position of the planet after each scan, a complete mapping of the arrays is performed. The frequency responses of LEKIDs are measured with a NIKEL readout system~\cite{Bourrion:2016a}, which is also used in the NIKA2 instrument.

\section{Results and Discussion}

\begin{figure}
\centering
\includegraphics[scale=0.36,trim={2cm 2cm 2cm 2cm},clip]{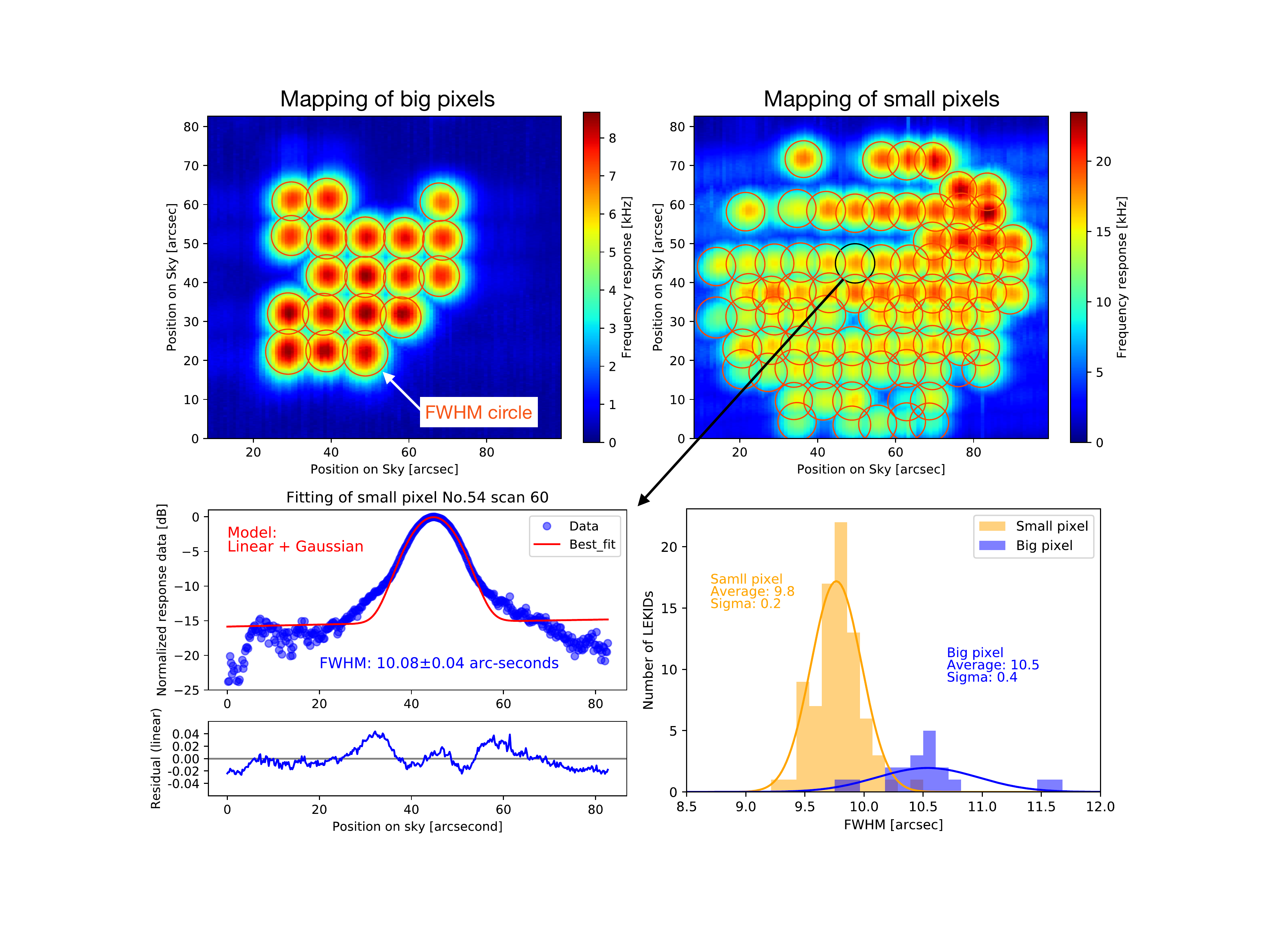}
\caption{\label{fig:map}Optical measurement results of the two designs. {\it Top} two panels show the mappings of two arrays. The {\it red} circles indicate the FWHM of the different pixels. {\it Bottom left} panel shows the fitting of one small pixel indicated by the {\it black} FWHM circle. The data shown is a scan in vertical direction, passing through the peak response and normalized to the maximum. {\it Bottom right} panel shows the histogram of resolutions for two arrays with Gaussian fit.}
\end{figure}

Two array mappings, shown in Fig.~\ref{fig:map}, are generated by combining all pixel responses together. The big pixel array has 25 pixels in design, of which 19 are mapped. The missing pixels are caused by the overlap of two resonances. The small pixel array contains 112 pixels, of which 82 are mapped. This low yield in small pixel array is mainly caused by the readout bandwidth of \SI{500}{MHz}. Due to a slight overestimation of the film kinetic inductance, the actual frequency range of the array is \SI{680}{MHz} instead of \SI{500}{MHz}.

The coordinates of the planet in the sky simulator are transferred to an on-sky angle by measuring the observed distance between two adjacent pixels and using the focal-plane-to-sky factor of \SI{4.875}{\arcsecond\per\milli\metre} for NIKA2 at the 30-m telescope. This calibration is not perfect, but the ratio between the resolutions of the two arrays is not affected by eventual errors in this calculation.

We used a 1-D Gaussian to fit the individual images of the planet, as shown in Fig.~\ref{fig:map}. This fit yields an average resolution of $9.8\pm 0.2\si{\arcsecond}$ and $10.5\pm 0.4\si{\arcsecond}$ for the small and big pixels, respectively, as illustrated in the histogram in Fig.~\ref{fig:map}. The large spread in the big pixel resolution is attributed to the limited number of 19 samples, and several outliers attributed to stray light reflected from the sample holder. The found increase of resolution of a factor $1.08\pm 0.05$ is consistent with the expected resolution increase of a factor $1.10$. With the factor $1.08$, we expect to achieve a resolution on sky of \SI{10.2}{\arcsecond} compared to the current value of \SI{10.9}{\arcsecond}.

\begin{table}[ht]
\caption{Characterization of the two arrays} 
\label{tab:parameters}
\begin{center}       
\begin{tabular}{l|M{1.3cm}|M{1.1cm}|M{1cm}|M{1.3cm}|M{1.7cm}|M{1.4cm}} 
\hline
\rule[-1ex]{0pt}{3.5ex}   & Inductor size [\si{mm^2}] & Inductor volume & Pitch [\si{mm^2}] & Resolution in-lab& Resolution on-sky & Responsivity [kHz] \\
\hline
\rule[-1ex]{0pt}{3.5ex}  Big pixel & $1.6\times 1.5$ & $\SI{1149}{\micro\metre^3}$ & $2\times 2$ & $10.5\pm 0.4\si{\arcsecond}$ & \SI{10.9}{\arcsecond}~\cite{Adam:2018a} & $106\pm 6$\\
\hline
\rule[-1ex]{0pt}{3.5ex}  Small pixel & $1.0\times 1.0$& $\SI{449}{\micro\metre^3}$ & $1.4\times 1.4$ & $9.8\pm 0.2\si{\arcsecond}$ & \SI{10.2}{\arcsecond}(expected) & $202\pm 32$\\
\hline
\end{tabular}
\end{center}
\end{table}

In Fig.~\ref{fig:map}, the nylon line, a point source, is transparent for big pixel design but our small pixels are quite sensitive to it, because of the \SI{60}{\percent} decrease in inductor volume. The final responsivity, shown in Table~\ref{tab:parameters} is extracted by the amplitude of the Gaussian fit showing the intensity of each response of the planet. The relative responsivity could be estimated with equation
\begin{equation}
Responsivity \propto \frac{\alpha Q}{V},
\end{equation}
where $\alpha$ is the kinetic inductance ratio, $Q$ is the loaded quality factor and $V$ is the inductor volume~\cite{Mazin:2005a}. In our case, the inductor volumes are \SI{449}{\micro\metre^3} and \SI{1149}{\micro\metre^3} and the loaded quality factors are about $8400$ and $12500$ for small and large pixel, respectively. Assuming the same kinetic inductance ratio, a factor of $1.7$ is derived compared to the observed factor $2$. The responsivities of pixels have a \SI{25}{\percent} non-uniformity on both arrays, which are caused by the uniformity of the inductor width. This non-uniformity changes the volume and sheet impedance of the inductor and the optical response is changed as the result~\cite{Shu:2018p}. For both two arrays in this measurement, there is no obvious difference in noise spectrum at frequencies larger than \SI{1}{Hz}, suggesting an improvement by a factor $2$ of sensitivity using the new pixel design. 

\section{Conclusion}

Table~\ref{tab:parameters} summarizes the results obtained in this study. We have successfully designed and characterized a compact pixel prototype for the \SI{260}{GHz} band of the NIKA2 instrument, that is expected to decrease the on-sky resolution from the actual \SI{10.9}{\arcsecond} to \SI{10.2}{\arcsecond}, while doubling the expected sensitivity. This design can be used to fill the complete \SI{6.5}{\arcminute} field of view of the NIKA2 camera once the bandwidth of the readout electronics is increased as planned.

\begin{acknowledgements}
The authors thank Akira Endo for help performing the simulations, and Dominique Billon Pierron and Arnaud Barbier for help with the array fabrication. This work has been partially funded by the LabEx FOCUS ANR-11-LABX-0013. 

\end{acknowledgements}

%\pagebreak

\bibliographystyle{JLTPv2}
\bibliography{Shu_lib.bib}
\end{document}